\documentclass[aps,groupedaddress,showpacs,floatfix]{revtex4}
\usepackage{amssymb}
\usepackage{amsmath}
\usepackage{graphicx}
\begin{document}

\title{Two-temperature statistics of free energies in (1+1) directed polymers}

\author{Victor Dotsenko}

\affiliation{LPTMC, Universit\'e Paris VI, Paris, France}

\affiliation{L.D.\ Landau Institute for Theoretical Physics, Moscow, Russia}

\date{\today}

\begin{abstract}
The joint statistical properties of two free energies
computed at two different temperatures  in {\it the same sample} 
of $(1+1)$ directed polymers is studied in terms of the replica technique.
The scaling dependence of the reduced free energies difference 
${\cal F} = F(T_{1})/T_{1} - F(T_{2})/T_{2}$ 
on the two temperatures $T_{1}$ and $T_{2}$ is derived. 
In particular, it is shown that if the two temperatures $T_{1} \, < \, T_{2}$
are close to each other the typical value of the fluctuating part of the  
reduced free energies difference ${\cal F}$ is proportional to $(1 - T_{1}/T_{2})^{1/3}$. 
It is also shown that the left tail asymptotics of this free energy difference probability
distribution function coincides with the corresponding tail of the TW distribution.
\end{abstract}


\maketitle

\medskip

\section{Introduction}

In this paper we consider the model of one-dimensional directed polymers in terms of an elastic string $\phi(\tau)$
directed along the $\tau$-axes within an interval $[0,t]$ which passes through a random medium
described by a random potential $V(\phi,\tau)$. This model is defined in terms of the Hamiltonian
\begin{equation}
   \label{1}
   H[\phi;\, V] = \int_{0}^{t} d\tau
   \Bigl[\frac{1}{2} \bigl[\partial_\tau \phi(\tau)\bigr]^2
   + V[\phi(\tau),\tau]\Bigr];
\end{equation}
where the disorder potential $V[\phi,\tau]$ is Gaussian distributed with a zero mean $\overline{V(\phi,\tau)}=0$
and the $\delta$-correlations
\begin{equation}
\label{2}
 {\overline{V(\phi,\tau)V(\phi',\tau')}} = u \delta(\tau-\tau') \delta(\phi-\phi') \,  .
\end{equation}
with the parameter $u$ describing the strength of the disorder.

This problem, which is equivalent to the one of the Kardar-Parisi-Zhang (KPZ) equation
\cite{KPZ} describing the time evolution of an interface
in the presence of noise, has been the focus  of intense studies during past three
decades
\cite{hh_zhang_95,burgers_74,kardar_book,hhf_85,numer1,numer2,kardar_87,bouchaud-orland,Brunet-Derrida,
Johansson,Prahofer-Spohn,Ferrari-Spohn1,KPZ-TW1a,KPZ-TW1b,KPZ-TW1c,KPZ-TW2,BA-TW2,BA-TW3,
LeDoussal1,LeDoussal2,goe,end-point,LeDoussal3,Corwin,Borodin}.
At present it is well established that depending on the boundary conditions 
the fluctuations of the free energy of the model defined by the
Hamiltonian (\ref{1}) are described by the GUE \cite{KPZ-TW1a,KPZ-TW1b,KPZ-TW1c,KPZ-TW2,BA-TW2,BA-TW3,LeDoussal1}, 
GOE \cite{LeDoussal2,goe} or GSE \cite{LeDoussal3}
Tracy-Widom  distribution \cite{TW-GUE}.
The two-point as well as $N$-point free energy distribution function
which describes joint statistics of the free energies of the directed polymers
coming to different endpoints has been derived in  \cite{Prolhac-Spohn,2pointPDF,Imamura-Sasamoto-Spohn,N-point-1,N-point-2}.
Besides, the joint statistical properties of the free energies at  
two different times has been studied in  
\cite{2time-1,2time-2,2time-J,Ferrari-Spohn_time-corr,Takeuchi_time-corr,2time-3,Nardis-LeDoussal-Takeuchi_time-corr}.

In the present paper I would like to propose one more "direction" of the studies of this system, namely, 
joint statistics of the free energies (or the interfaces, in the KPZ-language) at two different temperatures 
defined for the same quenched disorder. In other words, I am going to study the joint probability distribution function
of the free energies at two (or more) different temperatures for a given realization of the disorder potential $V[\phi,\tau]$.
Some years ago similar kind  of problem (under the name "temperature chaos") has been investigated for spin-glass-like
systems \cite{SG-T-chaos1,SG-T-chaos2,SG-T-chaos3,SG-T-chaos4} 
as well as for directed polymers on a hierarchical lattice \cite{DP-T-chaos}.
In this paper in terms of the standard replica formalism 
I derive  the general scaling dependence of the difference 
of two free energies  at two different temperatures, eqs.(\ref{35}) and (\ref{30}) as well as the 
left tail asymptotics of the corresponding universal probability distribution function, eq.(\ref{37}).  
In particular, it will be shown that if the two temperatures $T_{1} \, < \, T_{2}$ are close to each other,
so that $(1-T_{1}/T_{2}) \; \ll \; 1$, the difference
of the two free energies scales as $(1-T_{1}/T_{2})^{1/3} \, t^{1/3}$, eq.(\ref{42}).

\section{Replica Formalism}

For a fixed boundary conditions, $\phi(0) = \phi(t) = 0$, and for a given realization of disorder 
the partition function of the model defined in eqs.(\ref{1})-(\ref{2}) is
\begin{equation}
\label{3}
   Z(\beta, t) = \int_{\phi(0)=0}^{\phi(t)=0}
              {\cal D} \phi(\tau)  \;  \mbox{\Large e}^{-\beta H[\phi; \, V]}
\; = \; \exp\bigl(-\beta F(\beta, t)\bigr)
\end{equation}
where $\beta$ is the inverse temperature and $F(\beta, t)$ is the (random) free energy.
It is well known that in the limit $t\to\infty$ this free energy scales as
\begin{equation}
\label{4}
F(\beta, t) = f_{0}(\beta) \, t +  \frac{1}{2} \, (\beta u)^{2/3} \, t^{1/3} \, f \, ,
\end{equation}
where $f_{0}(\beta)$ is the (non-random) selfaveraging free energy density, and $f$
is a random quantity described by the
Tracy-Widom distribution.

For a given realization of the disorder potential $V[\phi,\tau]$ let us consider the above system at two
different temperatures $T_{1} \not= T_{2}$. More specifically, we are going to study
how the two free energies $F(\beta_{1}, t)$ and $F(\beta_{2}, t)$
of the same system are related to each other. In the present paper we are going to study
the statistical and scaling properties of the quantity
\begin{equation}
\label{5}
{\cal F}(\beta_{1}, \beta_{2}; \, t) \; = \;
\beta_{1} F(\beta_{1}, t) \, - \, \beta_{2} F(\beta_{2}, t)
\end{equation}
where, in what follows it will be assumed that $\beta_{1} \, > \, \beta_{2}$ (or $T_{1} \, < \, T_{2}$).
According to the definition (\ref{3})
\begin{equation}
\label{6}
\exp\bigl\{- {\cal F}(\beta_{1}, \beta_{2}; \, t) \bigr\} \; = \;
Z(\beta_{1}, t) \, Z^{-1}(\beta_{2}, t)
\end{equation}
Taking $N$-th power of the the both sides of the above relation and averaging over the disorder
we get
\begin{equation}
\label{7}
\int  d{\cal F}  P_{\beta_{1},\beta_{2},t}({\cal F})  \exp\bigl\{- N {\cal F}\bigr\}  = 
\overline{Z^{N}(\beta_{1}, t)  Z^{-N}(\beta_{2}, t)}
\end{equation}
where $\overline{(...)}$ denotes the averaging over the random potential $V$ and
$P_{\beta_{1},\beta_{2},t}({\cal F})$
is the probability distribution function of the random quantity ${\cal F}$, eq.(\ref{5}).
Introducing the replica partition function
\begin{equation}
\label{8}
{\cal Z}(M,N; \, \beta_{1}, \beta_{2}; \, t) \; = \;
\overline{Z^{N}(\beta_{1}, t) \, Z^{M-N}(\beta_{2}, t)}
\end{equation}
the relation (\ref{7}) can be formally represented as
\begin{equation}
\label{9}
\int  d{\cal F}  P_{\beta_{1},\beta_{2},t}({\cal F})  \exp\bigl\{- N {\cal F}\bigr\}  = 
\lim_{M\to 0} {\cal Z}(M,N;  \beta_{1}, \beta_{2};  t) \, .
\end{equation}
Following the standard "logic" of the replica technique, first it will be assumed that
both $M$ and $N$ are integers such that $M > N$. Next, after computing the replica partition
function ${\cal Z}(M,N; \, \beta_{1}, \beta_{2}; \, t)$ an analytic continuation for
arbitrary (complex) values of the parameters $M$ and $N$ has to be performed and the
limit $M \to 0$ has to be taken. After that, the relation (\ref{9}) can be considered
as the Laplace transform of the the probability distribution function
$P_{\beta_{1},\beta_{2},t}({\cal F})$ over the parameter $N$. In the case the function
${\cal Z}(0,N; \, \beta_{1}, \beta_{2}; \, t)$ had "good" analytic properties
in the complex plane of the argument $N$, this relation, at least formally, would allow
to reconstruct by inverse Laplace transform the probability distribution function
$P_{\beta_{1},\beta_{2},t}({\cal F})$. At present, for the considered problem
it is possible to derive an explicit expression for the function
${\cal Z}(0,N; \, \beta_{1}, \beta_{2}; \, t)$ only in the limit $N \gg 1$. Nevertheless,
using the relation (\ref{9}) this allows to reconstruct the left tail (${\cal F} \to -\infty$)
of the distribution function $P_{\beta_{1},\beta_{2},t}({\cal F})$. Moreover, it also allows to
derive the scaling dependence of free energy difference ${\cal F}$ on $\beta_{1},\beta_{2}$ and $t$.
Indeed, in the case in which the replica partition function has an exponential asymptotics
\begin{equation}
\label{10}
{\cal Z}(0,N \to \infty ;  \beta_{1}, \beta_{2};  t\to\infty)  \sim 
\exp\bigl\{A(\beta_{1}, \beta_{2})  t  N^{\alpha}\bigr\}  ,
\end{equation}
the left tail of the probability distribution function assumes the stretched-exponential form
\begin{equation}
\label{11}
P_{\beta_{1},\beta_{2},t}({\cal F}\to -\infty) \; \sim \;
\exp\bigl\{-B(\beta_{1}, \beta_{2}; \, t) \, |{\cal F}|^{\omega}\bigr\} \, .
\end{equation}
Then the saddle-point estimate of the integral on the l.h.s of eq.(\ref{9}) yields:
\begin{eqnarray}
\nonumber
&&\int  d{\cal F}  \exp\bigl\{-B\, |{\cal F}|^{\omega}+ N |{\cal F}|\bigr\} 
\\
\nonumber
\\
\nonumber
&&
\sim 
\exp\bigl\{(\omega-1)  \omega^{-\frac{\omega}{\omega-1}} B^{-\frac{1}{\omega-1}}
 N^{\frac{\omega}{\omega-1}}\bigr\} 
\\
\nonumber
\\
&&
\sim 
\exp\bigl\{A \, t \, N^{\alpha}\bigr\}
\label{12}
\end{eqnarray}
From this relation we find that
\begin{equation}
\label{13}
\omega \, = \, \alpha/(\alpha-1)
\end{equation}
and
\begin{equation}
\label{14}
B \; = \; (\alpha - 1) \, \alpha^{-\frac{\alpha}{\alpha-1}} \, \bigl(A \, t\bigr)^{-\frac{1}{\alpha-1}}
\end{equation}
Substituting this into eq.(\ref{11}) we get
\begin{equation}
\label{15}
P_{\beta_{1},\beta_{2},t}({\cal F}\to -\infty) \sim
\exp\Bigl\{
-(\alpha-1) \,
\Biggl(\frac{|{\cal F|}}{\alpha \bigl(A t \bigr)^{1/\alpha} }\Biggr)^{\frac{\alpha}{\alpha-1}}
\Bigr\} .
\end{equation}
If we assume that the (unknown) entire probability distribution function has a universal shape
the above asymptotic behavior implies that the considered quantity ${\cal F}$ scales as follows
\begin{equation}
\label{16}
{\cal F} \; = \; \bigl(A(\beta_{1}, \beta_{2})\bigr)^{1/\alpha} t^{1/\alpha} \, f
\end{equation}
where the random quantity $f\sim 1$ is described by some (unknown) probability distribution
function ${\cal P}(f)$ with the left asymptotics
${\cal P}(f\to -\infty)\sim \exp\bigl\{- (\mbox{const}) |f|^{\alpha/(\alpha - 1)}\bigr\}$.

Thus, the above speculations demonstrates that even if we know the replica partition function only in the limit $N \gg 1$,
we can still derive not only the left tail of the distribution function, but also (supposing that the entire 
distribution function is universal) the general scaling of the free energy.
In the next section we will demonstrate how this replica scheme can be applied for the concrete system under consideration.

\section{Mapping to Quantum Bosons}

According to eqs.(\ref{1})-(\ref{3}) and (\ref{8}), after performing the averaging over the random potential
we get
\begin{equation}
\label{17}
 {\cal Z}(M,N;\beta_{1},\beta_{2}; t)=
\prod_{a=1}^{M} \int_{\phi_{a}(0)=0}^{\phi_{a}(t)=0} {\cal D}\phi_{a}(\tau) 
\exp\bigl\{-H_{M}[\boldsymbol{\phi}]\bigr\}
\end{equation}
where $H_{M}[\boldsymbol{\phi}]$ is the replica Hamiltonian
\begin{eqnarray}
\nonumber
H_{M}[\boldsymbol{\phi}] &=&
\int_{0}^{t} d\tau
   \Biggl[
\frac{1}{2} \sum_{a=1}^{M} \beta_{a} \Bigl(\partial_\tau \phi_{a}(\tau)\Bigr)^2
\\
\nonumber
\\ 
&-&
\frac{1}{2} \, u^{2} \sum_{a\not b=1}^{M} \beta_{a} \beta_{b} \, \delta(\phi_{a} - \phi_{b})
\Biggr]
\label{18}
\end{eqnarray}
and
\begin{equation}
\label{19}
\beta_{a} \; = \;
\left\{
                          \begin{array}{ll}
\beta_{1}\,
\; \;
\mbox{for} \; a = 1, ..., N
\\
\\
\beta_{2}  \,
\; \;
\mbox{for} \; a = N+1, ..., M,
                          \end{array}
\right.
\end{equation}
Introducing:
\begin{equation}
\label{20}
\Psi(x_{1}, ..., x_{M} ; t) \; \equiv \;
\prod_{a=1}^{M} \int_{\phi_{a}(0)=0}^{\phi_{a}(t)=x_{a}} {\cal D}\phi_{a}(\tau) \;
\exp\bigl\{-H_{M}[\boldsymbol{\phi}]\bigr\}
\end{equation}
one can easily show that $\Psi({\bf x}; t)$ is the wave function of $M$-particle
boson system with attractive $\delta$-interaction defined by the Schr\"odinger equation:
\begin{eqnarray}
\nonumber
 -\partial_t \Psi({\bf x}; t) &=&
\sum_{a=1}^{M} \frac{1}{2\beta_{a}} \partial_{x_a}^2 \Psi({\bf x}; t) 
\\
\nonumber
\\ 
&+&
\frac{1}{2} u^{2} \sum_{a\not=b}^{M}\beta_{a}\beta_{b}\delta(x_{a}-x_{b}) \Psi({\bf x}; t)
\label{21}
\end{eqnarray}
with the initial condition $\Psi({\bf x}; 0) = \Pi_{a=1}^{M} \delta(x_a)$.
According to the definitions (\ref{17}) and (\ref{20}),
\begin{equation}
\label{22}
 {\cal Z}(M,N; \, \beta_{1}, \beta_{2}; \, t) \, = \,
 \Psi(x_{1}, ..., x_{M} ; t)\Big|_{x_{a}=0}
\end{equation}
The time dependent wave function  $\Psi({\bf x}; \, t)$ of the above quantum problem can be represented
in terms of the linear combination of the eigenfunctions $\Psi({\bf x})$ defined by the solutions of the
eigenvalue equation
\begin{equation}
   \label{23}
2E\Psi({\bf x}) =
\sum_{a=1}^{M} \frac{1}{\beta_{a}}\partial_{x_a}^2 \Psi({\bf x}) + 
 u^{2} \sum_{a\not=b}^{M} \beta_{a}\beta_{b} \delta(x_{a}-x_{b})\Psi({\bf x})
\end{equation}
Unlike the case with all $\beta$'s equal \cite{Lieb-Liniger,McGuire,Yang}, for the time being, the general solution of this equation
is not known. However, if we do not pretend to derive the exact result for the entire probability
distribution function $P_{\beta_{1},\beta_{2},t}({\cal F})$ but we want to know only its left tail
asymptotics  in the limit $t \to \infty$ then it would be enough
to get the behavior of the replica partition function
${\cal Z}(0,N \to \infty; \, \beta_{1}, \beta_{2}; \, t \to \infty)$
which is defined by the ground state solution only:
\begin{equation}
\label{24}
\Psi({\bf x}; t\to \infty) \; \sim \; \exp\bigl\{ - E_{g.s.}t\bigr\} \,  \Psi_{g.s.}({\bf x})
\end{equation}
One can easily check that the ground state solution of eq.(\ref{23}) is given by the eigenfunction
\begin{equation}
 \label{25}
\Psi_{g.s.}({\bf x}) \; \propto \;
\exp\Biggl\{
-\frac{1}{2} \, u \sum_{a,b=1}^{M} \gamma_{ab} \, \big|x_{a} - x_{b}\big|
\Biggr\}
\end{equation}
where
\begin{equation}
 \label{26}
\gamma_{ab} \; = \; \frac{\beta_{a}^{2} \, \beta_{b}^{2}}{\beta_{a} + \beta_{b}}
\end{equation}
The corresponding ground state energy is
\begin{equation}
 \label{27}
E_{g.s.}(M, N,\beta_{1},\beta_{2}) = -\frac{1}{2} u^{2}
\sum_{a=1}^{M} \frac{1}{\beta_{a}} \Biggl(\sum_{b=1}^{a-1}\gamma_{ab} - \sum_{b=a+1}^{M}\gamma_{ab}\Biggr)^{2}
\end{equation}
Note that in the trivial case $\beta_{1} = \beta_{2} = \beta$, using eqs.(\ref{25})-(\ref{27}),  one easily recovers the well known ground state solution
$\psi_{g.s.} \propto \exp\bigl\{ -\frac{1}{4} \, u \beta^{3} \sum_{a,b=1}^{M} \big|x_{a} - x_{b}\big| \bigr\}$
and $E_{g.s.} \; = \; -\frac{1}{24} u^{2}\beta^{5} (M^{3} - M)$.
Substituting eqs.(\ref{19}) and (\ref{26}) into eq.(\ref{27}) after simple algebra in the limit $M\to 0$ we obtain
\begin{equation}
 \label{29}
E_{g.s.}(0, N,\beta_{1},\beta_{2}) = -\frac{u^{2}}{24} \lambda(\beta_{1},\beta_{2}) N^{3}
+ \frac{u^{2}}{24}\bigl(\beta_{1}^{5}-\beta_{2}^{5}\bigr) \, N
\end{equation}
where
\begin{eqnarray}
 \nonumber
\lambda(\beta_{1},\beta_{2}) &=& 
4\bigl(\beta_{1}^{5}-\beta_{2}^{5}\bigr) 
- 
6 \bigl(\beta_{1}-\beta_{2}\bigr) 
\frac{2\beta_{1}^{4}\beta_{2}+2\beta_{2}^{4}\beta_{1} +\beta_{1}^{5} +\beta_{2}^{5}}{\beta_{1}+\beta_{2}} 
\\
\nonumber
\\
&+&
3 \bigl(\beta_{1}-\beta_{2}\bigr)^{2} 
\frac{\beta_{1}^{3}(2\beta_{2}+\beta_{1})^{2} - \beta_{2}^{3}(2\beta_{1}+\beta_{2})^{2}}{(\beta_{1}+\beta_{2})^{2}}
 \label{30}
\end{eqnarray}
According to eqs.(\ref{22}) and (\ref{24}) we find
\begin{eqnarray}
 \nonumber
&&{\cal Z}(0,N\to\infty;\beta_{1},\beta_{2}; t\to\infty) \sim
\\
\nonumber
\\
&&
\sim 
\exp\Bigl\{
\frac{u^{2}}{24}  \lambda(\beta_{1},\beta_{2}) \, N^{3}  t  
- 
\frac{ u^{2}}{24} \bigl(\beta_{1}^{5} - \beta_{2}^{5}\bigr)  N  t
\Bigr\}
 \label{31}
\end{eqnarray}
The second (linear in $N$ term) in the exponential of the above relation provides  the contribution to
the selfaveraging (non-random) linear in time part of the free energy variance ${\cal F}$.
Substituting eq.(\ref{31}) into eq.(\ref{9}) and redefining
\begin{equation}
 \label{32}
{\cal F} \; = \; \frac{1}{24} u^{2} \, \bigl(\beta_{1}^{5} - \beta_{2}^{5}\bigr) \,  t \; + \; \tilde{{\cal F}}
\end{equation}
we find that in the limits $t\to\infty$ and $N\to\infty$ the left tail of the probability distribution function
for the random quantity $\tilde{{\cal F}}$ (as $\tilde{{\cal F}} \to -\infty$) is defined by the relation
\begin{equation}
\label{33}
\int  d\tilde{{\cal F}} \, P_{\beta_{1},\beta_{2},t}(\tilde{{\cal F}})  \exp\bigl\{- N \tilde{{\cal F}} \bigr\} \sim
\exp\Bigl\{
\frac{u^{2}}{24}  \, \lambda(\beta_{1},\beta_{2})  N^{3}  t
\Bigr\} .
\end{equation}
Redefining
\begin{equation}
\label{34}
N \; = \; 2 (u^{2} \lambda)^{-1/3} \, s
\end{equation}
we find that the free energy difference $\tilde{{\cal F}}$ scales as
\begin{equation}
\label{35}
\tilde{{\cal F}} \; = \; 
\frac{1}{2} \, u^{2/3} \, \bigl(\lambda(\beta_{1},\beta_{2})\bigr)^{1/3} \, t^{1/3} \, f
\end{equation}
where the left tail of the {\it universal} probability distribution function ${\cal P}(f)$ 
of the random quantity $f$
is defined by the relation
\begin{equation}
\label{36}
\int \, df \, {\cal P}(f) \, \exp\bigl\{- s \, f \bigr\} \; \sim \;
\exp\Bigl\{ \frac{1}{3} \, s^{3} \Bigr\} .
\end{equation}
A simple saddle-point estimate of the above integral (for $s \gg 1$ and $|f| \gg 1$) yields
\begin{equation}
\label{37}
{\cal P}(f \to -\infty) \; \sim \; 
\exp\Bigl\{ -\frac{2}{3} \, |f|^{2/3} \Bigr\} .
\end{equation}
Note that this tail coincides with the corresponding asymptotics 
of the usual free energy TW distribution \cite{TW-GUE}.

\vspace{5mm}

Let us consider in more detail the scaling relation (\ref{35}) which demonstrate the dependence
of the typical value of the fluctuating part of the reduced free energy difference, eq.(\ref{5}), 
on the strength of disorder $u$, on  the inverse temperatures $\beta_{1}$ and $\beta_{2}$,  
and on time $t$. First of all, one notes that the disorder scaling $\sim u^{2/3}$ as well as  
the time scaling $\sim t^{1/3}$ coincide with the ones  of the usual free energy scaling in 
$(1+1)$ directed polymers, which of course is not surprising. 
On the other hand, the dependence on the inverse temperatures $\beta_{1}$ and $\beta_{2}$ 
turns out to be   less trivial.

First of all, using explicit expression (\ref{30}) one easily finds that in the limit
$\beta_{1} \gg \beta_{2}$  (or $T_{1} \ll T_{2}$)
\begin{equation}
 \label{38}
\lambda\bigl(\beta_{1}, \beta_{2}\bigr)\big|_{\beta_{1}\gg \beta_{2}} \, \simeq \, \beta_{1}^{5} \; ,
\end{equation}
so that in this limit the scaling relation (\ref{35}) turns into the usual one-temperature
free energy scaling 
\begin{equation}
 \label{39}
\tilde{{\cal F}} \; \simeq \; \beta_{1} \tilde{F}_{1} \, = \, \frac{1}{2} \bigl(u^{2}\beta_{1}^{5}\bigr)^{1/3} \, t^{1/3} \, f
\end{equation}
In other words, in this case the free energy $F_{1}$ of the polymer with the temperature $T_{1}$ is much lower 
than that of the polymer with the temperature $T_{2} \gg T_{1}$, and the free energy difference
$\tilde{{\cal F}}$ is dominated by the free energy $F_{1}$ as it should be.

Let us consider now what happens if the two temperature parameters $\beta_{1}$ and $\beta_{2}$ are
close to each other. Introducing a small (positive) parameter 
\begin{equation}
 \label{40}
\epsilon \; = \; \frac{\beta_{1} \, - \, \beta_{2}}{\beta_{1}} \, \ll \, 1
\end{equation}
and substituting $\beta_{2} \, = \, (1-\epsilon) \beta_{1}$ into eq.(\ref{30}) in the leading order in $\epsilon \ll 1$
we get
\begin{equation}
 \label{41}
\lambda \; \simeq \; 2\beta_{1}^{5} \, \epsilon
\end{equation}
Substituting this into eq.(\ref{35}) we find that in this case   the fluctuating part of the 
the corresponding free energy difference $\tilde{{\cal F}}$, eq.(\ref{5}), scales as
\begin{equation}
\label{42}
\tilde{{\cal F}}\; \simeq \; 
\frac{1}{2} \bigl(2 u^{2}\beta_{1}^{5}\bigr)^{1/3} \, 
\Biggl(\frac{\beta_{1} \, - \, \beta_{2}}{\beta_{1}}\Biggr)^{1/3} \, t^{1/3} \, f
\end{equation}
where the random quantity $f$ is described by a universal distribution function ${\cal P}(f)$ whose
left tail asymptotics is given in eq.(\ref{37}). 
The above eq.(\ref{42}) constitutes the main result of the present study.

\section{Conclusions}

In this paper we have studied the joint statistical properties of two  free energies
computed at two different temperatures  in {\it the  same sample} (i.e. for a given realization of the disorder)
of $(1+1)$ directed polymers. In particular, it is shown that if the two temperatures $T_{1} \, < T_{2}$ 
are close to each other the typical value of the fluctuating part of the reduced free energies difference 
${\cal F} = F(T_{1})/T_{1} - F(T_{2})/T_{2}$
is proportional to $(1 - T_{1}/T_{2})^{1/3}$, eq.(\ref{42}), which implies "one more 1/3"
exponent in these type of systems. On the other hand, the joint distribution function of these two free
energies for the time being remains unknown. The left tail of this free energy difference probability
distribution function, eq.(\ref{37}), coincides with corresponding tails of the TW distributions 
(both GUE, GOE and GSE) but this tells nothing about its entire exact shape. 

Unfortunately, in real experimental studies of the KPZ type systems, typically for a given realization of the disorder
the measurement of the statistical properties of the evolving interface profile (which by mapping to the directed polymers
corresponds to the free energy) can be done only once. In other words, each subsequent measurement implies  
a new realization of the disorder (see e.g. \cite{exper-rev}). 
For that reason, at present stage the possibility of a real experimental 
study of the effects considered in this paper looks rather problematic. 
On the other hand, the numerical investigation seems to be quite accessible. Compared with usual protocol of
the previous "one-temperature" studies (see e.g.\cite{Halpin-Healy} and references therein), 
one has just to repeat each measurement twice: keeping the same realization of the 
disorder but changing the temperature parameter to another value.


\begin{thebibliography}{0}
 




\bibitem{KPZ} M. \ Kardar, G.\ Parisi and Y-C \ Zhang, 
    Phys.\ Rev.\ Lett.\ {\bf 56}, 889 (1986)

\bibitem{hh_zhang_95} T.\ Halpin-Healy and Y-C.\ Zhang,
   Phys.\ Rep.\ {\bf 254}, 215 (1995)


\bibitem{burgers_74} J.M.\ Burgers, {\it The Nonlinear
   Diffusion Equation} (Reidel, Dordrecht, (1974))

\bibitem{kardar_book} M.\ Kardar,
   {\it Statistical physics of fields} (Cambridge: Cambridge University Press, (2007))


\bibitem{hhf_85} D.A.\ Huse, C.L.\ Henley, and D.S.\ Fisher,
    Phys.\ Rev.\ Lett.\ {\bf 55}, 2924 (1985)

\bibitem{numer1} D.A.\ Huse and C.L.\ Henley,
    Phys.\ Rev.\ Lett. {\bf 54}, 2708 (1985)

\bibitem{numer2} M.\ Kardar and Y-C.\ Zhang,
    Phys.\ Rev.\ Lett.\  {\bf 58}, 2087 (1987)


\bibitem{kardar_87} M.\ Kardar,
   Nucl.\ Phys.\ {\bf B 290}, 582 (1987)


\bibitem{bouchaud-orland} J.\ P.\ Bouchaud and H.\ Orland,
   J.\ Stat.\ Phys.\ {\bf 61}, 877 (1990)

\bibitem{Brunet-Derrida} E.\ Brunet and B.\ Derrida,
   Phys. \ Rev. \ E {\bf 61}, 6789 (2000)

\bibitem{Johansson}  K.\ Johansson,
   Comm. \ Math. \ Phys. \ {\bf 209}, 437 (2000)

\bibitem{Prahofer-Spohn} M.\ Prahofer and H.\ Spohn,
   J.\ Stat.\ Phys.\ {\bf 108}, 1071 (2002)

\bibitem{Ferrari-Spohn1} P.\ L.\ Ferrari and H.\ Spohn,
   Comm. \ Math. \ Phys. \ {\bf 265}, 1 (2006)


\bibitem{KPZ-TW1a} T.\ Sasamoto and H.\ Spohn,
         Phys.\ Rev.\ Lett.\ {\bf 104}, 230602 (2010)

\bibitem{KPZ-TW1b} T.Sasamoto and H.Spohn,
         Nucl.\ Phys.\ {\bf B834}, 523 (2010)

\bibitem{KPZ-TW1c} T.\ Sasamoto and H.\ Spohn,
         J.\ Stat.\ Phys. {\bf 140}, 209 (2010)

\bibitem{KPZ-TW2}  G.\ Amir, I.\ Corwin and J.\ Quastel,
        Comm.\ Pure Appl.\ Math.\ {\bf 64}, 466 (2011)


\bibitem{BA-TW2} V.\ Dotsenko,
         EPL, {\bf 90},20003 (2010)

\bibitem{LeDoussal1} P.\ Calabrese, P. Le Doussal and A.\ Rosso,
         EPL, {\bf 90},20002 (2010)

\bibitem{BA-TW3}  V.\ Dotsenko,
         J.Stat.Mech. P07010 (2010)

\bibitem{LeDoussal2} P.\ Calabrese and P. Le Doussal,
         Phys.\ Rev.\ Lett.\ {\bf 106}, 250603 (2011)


\bibitem{goe} V.\ Dotsenko,
         J. Stat. Mech. P11014 (2012)

\bibitem{end-point} V.\ Dotsenko,
         J. Stat. Mech. P02012 (2012)

\bibitem{LeDoussal3} T.Gueudr\'e and P. Le Doussal,
        EPL, {\bf 100}, 26006 (2012)

 \bibitem{Corwin} I.\ Corwin,
       Random Matrices: Theory Appl. {\bf 1}, 1130001 (2012)


\bibitem{Borodin} A.\ Borodin, I.\ Corwin and P.\ Ferrari,
        Comm. Pure Appl. Math. {\bf 67}, 1129–1214 (2014)
       


\bibitem{TW-GUE} C.A.\ Tracy and H.\ Widom,
         Commun.Math Phys. {\bf 159}, 151 (1994)





\bibitem{Prolhac-Spohn} S.\ Prolhac and H.\ Spohn,
         J.Stat.Mech. P01031 (2011)

\bibitem{2pointPDF} V.\ Dotsenko, 
        J.Phys. A: Math. Theor. {\bf 46}, 355001 (2013)

\bibitem{Imamura-Sasamoto-Spohn} T.\ Imamura, T.\ Sasamoto and H.\ Spohn,
        J.Phys. A: Math. Theor. {bf 46}, 355002 (2013)


\bibitem{N-point-1} S.\ Prolhac and H.\ Spohn,
        J.Stat.Mech. P03020 (2011)

\bibitem{N-point-2} V.\ Dotsenko, 
        Cond.Mat.Phys. {\bf 17}, 33003 (2014)


\bibitem{2time-1} V.\ Dotsenko, 
    J.Stat.Mech. P06017 (2013)

\bibitem{2time-2} V.\ Dotsenko,
    J.Phys. A: Math. Theor. {\bf 48} 495001 (2015)

\bibitem{2time-J} K.\ Johansson, 
    arXiv:1502.00941 (2015)

\bibitem{Ferrari-Spohn_time-corr} P.L.\ Ferrari and H.\ Spohn, 
     SIGMA {\bf 12}, 074 (2016)
    
\bibitem{Takeuchi_time-corr} K.A.\ Takeuchi and M.\ Sano,
    J. Stat. Phys. {\bf 147}, 853-890 (2012)



\bibitem{2time-3} V.\ Dotsenko,
    J.Phys. A: Math. Theor. {\bf 49} 27LT01 (2016)


\bibitem{Nardis-LeDoussal-Takeuchi_time-corr} J.\ De Nardis, P.\ Le Doussal and K.A.\ Takeushi,
    arXiv:1611.04756v1 (2016)



\bibitem{SG-T-chaos1} T.\ Rizzo,
   Eur. Phys. J. {\bf B 29}, 425 (2002)


\bibitem{SG-T-chaos2} T. Rizzo and A.\ Crisanti
     Phys. Rev. Lett. {\bf 90}, 137201 (2003)

\bibitem{SG-T-chaos3} F.\ Krzakala  and J.-P.\ Bouchaud, arXiv:cond-mat/0507555v2 (2005)

\bibitem{SG-T-chaos4} P.\ Barucca, G.\  Parisi and T.\ Rizzo,
     Phys. Rev. {\bf E 89}, 032129 (2014)

\bibitem{DP-T-chaos} R.\ A.\ Da Silveira and J.-P. Bouchaud,
      Phys. Rev. Lett.,  {\bf 93}, 015901 (2004)








\bibitem{Lieb-Liniger} E.H.\ Lieb and W.\ Liniger,
     Phys.\ Rev.\  {\bf 130}, 1605 (1963)

\bibitem{McGuire} J.B.\ McGuire,
     J.\ Math.\ Phys.\ {\bf 5}, 622 (1964).

\bibitem{Yang} C.N.\ Yang,
     Phys.\ Rev.\  {\bf 168}, 1920 (1968)







\bibitem{exper-rev} K.A.\ Takeuchi,
      J.Stat.Mech.: Th.Exp.,  P01006 (2014)

\bibitem{Halpin-Healy} T.\ Halpin-Healy, arXiv:1310.8013v1 (2013)


\end{thebibliography}
\end{document}